# Nouvelle méthode d'identification paramétrique d'un modèle de préférence

J. Renaud, M. Camargo[a], L. Morel[a], C. Fonteix[a]

[a]ERPI, Equipe de Recherche des processus Innovatifs, EA 3767
8, rue Bastien Lepage, 54010 Nancy Cedex, France

**Résumé**

Cet article présente une contribution à l'aide à la décision multicritère destinée aux décideurs industriels afin de déterminer le meilleur compromis entre les critères de conception lorsqu'ils travaillent sur des produits à risque ou innovants. Dans (RENAUD et al. 2008) nous utilisions l'opérateur OWA (Ordered Weighted Average), il s'agit d'une technique d'analyse multicritère bien connue introduite par (YAGER 1988). L'intérêt de cette méthode d'agrégation est, au-delà de sa simplicité d'utilisation, sa capacité d'évaluation d'un produit selon une échelle unique. Lorsqu'on utilise la méthode OWA, le choix des poids des critères et de leurs valeurs reste une opération importante et délicate. En effet, les poids ne sont pas fixés par critère mais selon un niveau d'utilité (FISHBURN 1967). Les poids peuvent, en outre, être déterminés à l'aide de différentes méthodes.

Une approche traditionnelle consiste à estimer les poids à partir d'une sélection a priori des échantillons les plus représentatifs par un expert. Nous proposons une nouvelle approche basée sur la D-Optimalité appliquée afin de déterminer le meilleur échantillon. Les résultats des deux approches sont comparés. En effet, dans certains problèmes de décision multicritères, il est difficile de choisir la méthode qui décrit le mieux le comportement du décideur. Comme dit plus haut, la première partie du présent article présente une méthode originale basée sur la D-optimalité pour déterminer l'ensemble d'échantillons le plus représentatif permettant de déterminer les paramètres de décision. Cette méthode a été appliquée et validée dans une approche OWA. Il a été démontré que la précision et la fiabilité de la méthode ont été améliorées.

L'application OWA ayant été améliorée, l'autre but du présent article est de vérifier si le vrai décideur a un comportement basé sur OWA ou non. Pour y parvenir, l'approche D-optimalité est appliquée à l'approche MAUT (théorie de l'utilité multi-attributs) afin de trouver le meilleur ensemble d'échantillons. Les résultats des deux méthodes sont comparés. Les deux approches montrent que, si OWA simule mieux le comportement du décideur, il y a toujours un écart entre ses notes et celle de OWA. La question ici est de savoir comment améliorer la précision de la note estimée par OWA par rapport à celles données par l'expert.

Par la suite, une approche hybride est testée, comme une combinaison linéaire des approches MAUT et OWA. Les résultats obtenus avec cette combinaison se sont révélés plus précis que ceux de l'OWA. Cependant, nous avons également testé l'approche de l'intégrale discrète de CHOQUET. Autrement dit, il est possible de trouver un modèle capable de mieux décrire le comportement du décideur.

Mots clé : analyse multicritère ; OWA ; poids des critères ; D-Optimalité

## 1 Introduction

L'objectif principal de l'application des techniques de prise de décision multicritère (MCDM) (VINCKE 1989) (RENAUD et al. 2007) dans le processus de développement de nouveaux produits est d'adapter au mieux les fonctionnalités du produit à concevoir aux besoins du client (PORTER 1984) (WU et al. 1997). Une fois les critères pertinents identifiés, le calcul des pondérations relatives à attribuer à chaque critère devient une préoccupation importante (FULLER et al. 2001) (XU 2004a).

Comme un critère ou une performance d'un produit est souvent plus important que les autres, d'un point de vue subjectif ou objectif, la différenciation entre eux se caractérise par le poids du critère ou de la performance du produit ($w_i$). S'il existe plusieurs critères, un poids est affecté à chacun d'eux, alors un vecteur de poids est défini comme : $\vec{w}^T = (w_1 \quad \cdots, w_i \quad \cdots, w_n)$ où le nombre de critères est $n$. La valeur des poids pouvant dépendre de la méthode d'agrégation choisie, la première étape consiste donc à définir la méthode d'analyse multicritère puis à déterminer le poids de chaque critère. Cependant, l'estimation des poids n'est pas une tâche facile, mais elle a été largement étudiée et a donné lieu à plusieurs travaux de recherche (ECKENRODE 1965) (NIJKAMP et al. 1981) (FARQUHAR et al. 1989). Il y a deux possibilités indiquées figure 1 :

1. Les principaux critères sont bien connus sur l'ensemble de la population, dans ce cas les poids pourraient être calculés comme suit :
   - le décideur est en mesure de proposer une hiérarchie des critères selon sa préférence. Une note est attribuée pour chaque critère. Les poids sont calculés par normalisation des scores (EDWARDS 1977). D'une autre manière, l'approche MACBETH propose une procédure de questionnement simple pour « piloter » la quantification interactive des valeurs de poids (BANA E COSTA et al. 1994).
   - le décideur n'est pas en mesure de donner une hiérarchie de critères. Il ou elle compare les critères d'égal à égal en considérant les différentes alternatives puis définit le niveau de priorité de chacune d'elles par rapport à chaque critère., et les poids sont calculés par la méthode AHP définie par (SAATY 1977).

2. Les poids des critères seront déterminés en étudiant un échantillon sur une population, en l'occurrence :
   - Le décideur n'est pas en mesure d'attribuer une pondération à chaque critère, mais il est en mesure d'attribuer une note à chaque produit d'un échantillon donné. Dans ce cas, les poids sont calculés par identification paramétrique.
   - Le décideur n'est pas en mesure de donner un poids à chaque critère, et il n'est pas en mesure de donner une note à chaque produit d'un échantillon donné, mais il est en mesure de classer les produits de l'échantillon. Dans ce cas, les poids peuvent être calculés par identification paramétrique avec l'hypothèse d'une répartition uniforme des scores. Cette méthode est développée dans (RENAUD et al. 2008).

Dans le tout dernier cas, les auteurs ont proposé une nouvelle approche d'identification des poids pour les opérateurs OWA (FILEV et al. 1998) (CARLSSON et al. 2003) (CHAKRABORTY et al. 2004). Leur méthodologie nécessite le choix d'un échantillon, classé par le décideur, utilisé pour la détermination des valeurs de pondération. Le principe de résolution est basé sur l'équipartition du score de chaque échantillon de produit. Trois échantillons ont été construits à l'aide des valeurs les plus fortes, les plus faibles et médianes de chaque critère, et un quatrième a été sélectionné avec l'assistance de l'expert. L'objectif principal du présent article est de comparer la technique précédente à la nouvelle, sur la base de l'attribution de scores pour chaque produit d'un échantillon donné par le décideur. Dans les deux cas le poids de chaque critère est calculé par identification paramétrique. Afin de déterminer la technique la plus efficace, nous comparons l'écart entre les poids identifiés et un ensemble de poids de référence. Ces poids de référence ne peuvent pas être ceux du véritable décideur, car ses préférences n'ont pu être déterminées précisément par aucune des techniques utilisées.

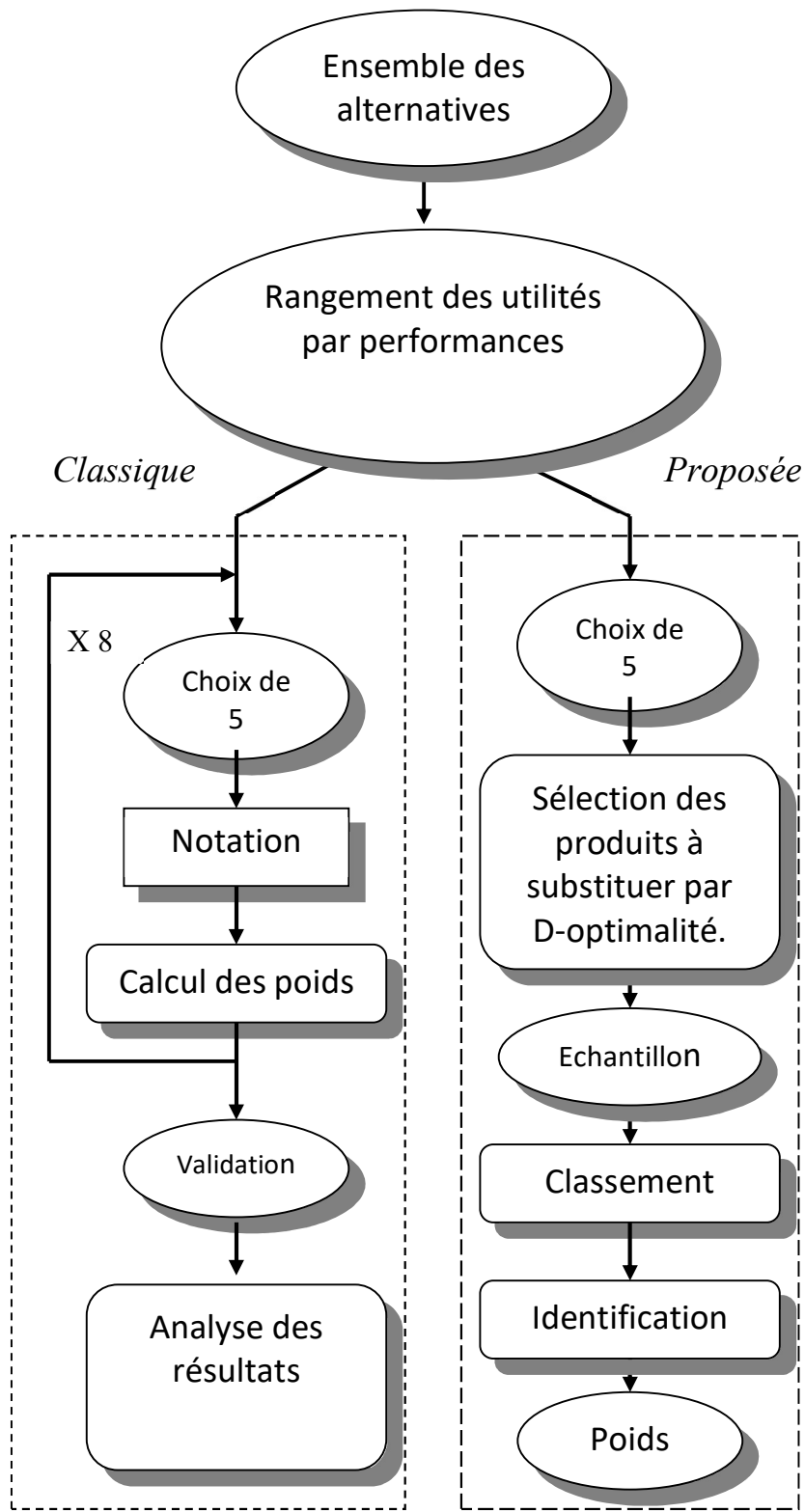

Figure 1 : Les deux processus d'identification paramétrique utilisés

Ainsi un décideur virtuel sera créé avec des valeurs de poids fixes permettant de calculer le score théorique de chaque produit. Afin de simuler le comportement réel du décideur, une erreur gaussienne centrée sera ajoutée au score théorique à valeur de variance connue. Trois valeurs de variance seront utilisées pour calculer les scores simulés permettant d'étudier l'effet de la valeur de cette variance sur la précision de l'identification.

Une application industrielle issue de l'agroalimentaire, notamment une production fromagère, illustre cette comparaison des techniques. Ces types de produits s'adressent à des consommateurs gourmets correspondant à un marché différencié. Parmi les méthodes multicritères (ZELENY 1982) (BRANS et al. 1984a) (BRANS et al. 1984b) (BOUYSSOU 1992) (ROY et al. 1993) (MAYSTRE et al. 1994), l'OWA (Ordered Weighted Average) a été retenue, car elle permet d'attribuer des pondérations aux performances et non à des critères bien définis (TORRA 1997) (YAGER 2003) (XU 2004b). Cela correspond au type de consommateur cité et a été validé par le véritable décideur. En effet, dans ce cas, l'objectif est de définir, mesurer et ajuster les propriétés de fromages haut de gamme évalués par un panel d'experts. Une technique de surclassement non compensatoire comme l'OWA permet de concevoir les fromages spéciaux préférés des consommateurs.

La section 2 présente l'application industrielle, les opérateurs OWA seront vu plus loin dans la section 3, les résultats sont discutés en section 4 et 5, et l'approche par D-optimalité est décrite en section 6. Après une validation et une analyse des résultats, une approche par combinaison linéaire OWA-MAUT est présentée en section 7 puis une agrégation des critères par l'intégrale de Choquet est effectuée dans la section 8, et pour finir par une conclusion dans la section 9.

## 2 Application industrielle

Le secteur concerné est l'industrie agroalimentaire, notamment un producteur de fromage haut de gamme de l'Est de la France. Plusieurs fois par mois, un groupe d'experts se réunit pour évaluer différents échantillons de la ligne de production. Un lot de 47 fromages est traité dans le cadre du présent article, représentant les mesures récupérées sur une période de trois mois. (Tableau 1).

Dix-huit critères ont été notés de 0 à 7 (en nombre décimal) pour chaque fromage et l'écart à une cible a été calculée pour chaque critère. Une moyenne a été déduite pour chaque critère, puis certains critères ont été agrégés pour qu'il n'en reste que quatre. Par exemple, le « grand » critère correspondant à l'aspect extérieur regroupe 5 sous critères. Il en est de même de la texture et du goût, ou arôme, par contre l'odeur, ou le parfum, n'en regroupe que trois. Ainsi nous obtenons 4 critères par fromage : l'aspect, noté C1, l'odeur, notée C2, la texture, notée C3 et le goût noté C4. L'écart à la cible est moyenné et transformé à l'aide d'une quantification floue présentée dans (RENAUD et al. 2008). La valeur prise en compte est donc comprise entre 0 et 1.

La qualité du fromage dépend des valeurs des conditions opératoires du procédé comme le temps de processus, le taux d'extraction, le débit d'écoulement de la cuve, le débit de transfert de cuve, etc. La figure 2 montre un schéma du procédé.
- Collecte du lait toutes les 48 heures en cuve (-4°C),
- Le lait est écrémé et acidifié par ajout d'agents de fermentation, puis il est pasteurisé,
- Pressage et découpe du futur fromage,
- Phase de décantation, pendant 20 heures, le fromage est retourné deux fois,
- Egouttage, pré-maturage et affinage.

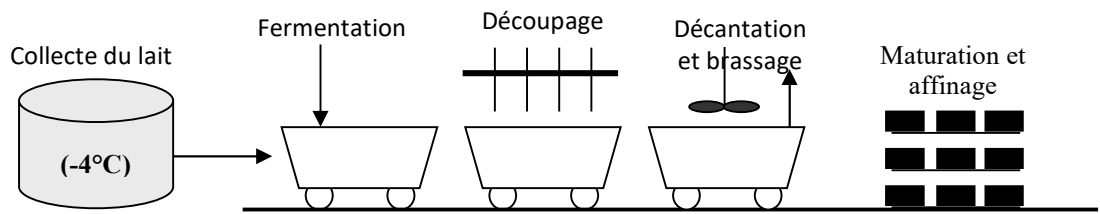

Figure 2 : Procédé de fabrication du fromage

## 3 Méthodologie

Nous avons dit que l'opérateur OWA convient mieux à un produit de niche, la particularité de cet opérateur c'est qu'il définit autant de poids qu'il y a de critère ($w_1$, $w_2$, $w_3$ et $w_4$) mais que ces poids ne sont pas affectés à un critère, mais une performance notée $j$ pour le critère $\sigma(j)$. Par exemple, le produit 1813 est tel que C1=0,465, C2=0,705, C3=0,700 et C4=0,946, donc $\sigma(1) = 4, \sigma(2) = 2, \sigma(3) = 3$ et $\sigma(4) = 1$, car le 4$^{\text{ème}}$ critère a la meilleure performance (0,946>0,705>0,700>0,465) et le premier est dernier.

| Produit | Critère 1 | Critère 2 | Critère 3 | Critère 4 | Scores | Note avec variance 0,1 | Note avec variance 0,5 | Note avec variance 1 |
|---|---|---|---|---|---|---|---|---|
| 162 | 0,774 | 0,824 | 0,734 | 0,000 | 0,672 | 5,12 | 5,23 | 4,48 |
| 292 | 0,814 | 0,970 | 0,400 | 0,338 | 0,743 | 7,37 | 6,55 | 7,31 |
| 592 | 0,930 | 0,794 | 0,800 | 0,676 | 0,832 | 10,50 | 9,64 | 10,15 |
| 612 | 0,698 | 0,441 | 0,800 | 0,676 | 0,697 | 6,02 | 5,55 | 6,39 |
| 613 | 0,621 | 0,705 | 0,734 | 0,992 | 0,816 | 10,13 | 10,03 | 9,26 |
| 703 | 0,814 | 0,353 | 0,700 | 0,909 | 0,766 | 8,08 | 8,66 | 6,95 |
| 733 | 0,233 | 0,705 | 0,800 | 0,811 | 0,705 | 5,20 | 6,92 | 6,09 |
| 813 | 0,814 | 0,705 | 0,700 | 0,909 | 0,819 | 9,52 | 8,91 | 9,67 |
| 823 | 0,930 | 0,353 | 0,400 | 0,606 | 0,667 | 5,06 | 5,13 | 7,04 |
| 873 | 0,814 | 0,441 | 0,700 | 0,606 | 0,693 | 5,39 | 5,49 | 6,34 |
| 963 | 0,814 | 0,705 | 0,466 | 0,809 | 0,744 | 7,02 | 7,74 | 6,71 |
| 1103 | 0,698 | 0,705 | 0,200 | 0,606 | 0,612 | 3,77 | 3,43 | 4,01 |
| 1193 | 0,814 | 0,794 | 0,400 | 0,909 | 0,787 | 8,63 | 8,08 | 8,52 |
| 1233 | 0,926 | 0,705 | 0,550 | 0,743 | 0,782 | 8,30 | 8,85 | 8,20 |
| 1293 | 0,930 | 0,441 | 0,600 | 0,541 | 0,699 | 6,17 | 4,09 | 3,96 |
| 1303 | 0,814 | 0,617 | 0,800 | 0,541 | 0,739 | 6,73 | 8,10 | 6,71 |
| 1323 | 0,930 | 0,529 | 0,900 | 0,811 | 0,843 | 10,52 | 9,98 | 9,77 |
| 1433 | 0,698 | 0,573 | 0,800 | 0,743 | 0,734 | 7,24 | 6,64 | 7,75 |
| 1513 | 0,774 | 0,529 | 0,600 | 0,541 | 0,650 | 3,94 | 5,08 | 3,24 |
| 1583 | 0,814 | 0,750 | 0,550 | 0,743 | 0,744 | 7,35 | 7,46 | 6,03 |
| 1793 | 0,741 | 0,705 | 0,800 | 0,541 | 0,729 | 6,54 | 7,74 | 6,56 |
| 1813 | 0,465 | 0,705 | 0,700 | 0,946 | 0,765 | 7,68 | 6,63 | 9,09 |
| 1853 | 0,741 | 0,705 | 1,000 | 0,606 | 0,819 | 9,75 | 11,17 | 9,90 |
| 1873 | 0,698 | 0,705 | 0,800 | 0,606 | 0,727 | 6,76 | 6,70 | 6,76 |
| 1923 | 0,741 | 0,587 | 0,666 | 0,809 | 0,734 | 7,27 | 6,58 | 7,23 |
| 2083 | 0,744 | 0,317 | 0,440 | 0,432 | 0,542 | 1,26 | 2,50 | 1,13 |
| 2343 | 0,930 | 0,750 | 0,650 | 0,811 | 0,825 | 9,85 | 10,88 | 9,73 |
| 2483 | 0,523 | 0,705 | 0,700 | 0,676 | 0,672 | 5,39 | 5,08 | 4,87 |
| 2553 | 0,658 | 0,353 | 0,400 | 0,765 | 0,616 | 3,36 | 3,72 | 3,37 |
| 2573 | 0,891 | 0,852 | 0,666 | 0,541 | 0,793 | 8,24 | 7,87 | 9,29 |
| 2583 | 0,542 | 0,529 | 0,734 | 0,586 | 0,630 | 4,14 | 3,01 | 2,03 |
| 2663 | 0,988 | 0,309 | 0,850 | 0,526 | 0,768 | 7,85 | 8,20 | 7,17 |
| 2693 | 0,97 | 0,1 | 0,865 | 0,473 | 0,741 | 7,09 | 6.66 | 9,17 |
| 2902 | 0,698 | 0,838 | 0,600 | 0,473 | 0,705 | 6,14 | 5,03 | 8,61 |
| 2922 | 0,833 | 0,750 | 0,900 | 0,473 | 0,793 | 8,61 | 9,07 | 8,37 |
| 2972 | 0,863 | 0,529 | 0,266 | 0,630 | 0,653 | 4,43 | 5,24 | 3,28 |
| 3062 | 0,581 | 0,176 | 0,700 | 0,676 | 0,596 | 3,11 | 2,71 | 3,61 |
| 3072 | 0,814 | 0,529 | 0,400 | 0,270 | 0,585 | 2,71 | 1,43 | 3,78 |
| 3112 | 0,640 | 0,794 | 0,550 | 0,541 | 0,673 | 5,61 | 4,30 | 7,31 |
| 3162 | 0,407 | 0,309 | 0,350 | 0,946 | 0,599 | 2,77 | 3,00 | 2,64 |
| 3192 | 0,698 | 0,353 | 0,600 | 0,405 | 0,573 | 2,67 | 2,00 | 1,96 |
| 3262 | 0,698 | 0,459 | 0,560 | 0,486 | 0,589 | 3,10 | 2,19 | 3,07 |
| 3302 | 0,930 | 0,529 | 0,560 | 0,703 | 0,746 | 7,02 | 7,84 | 7,97 |
| 3322 | 0,814 | 0,794 | 0,450 | 0,405 | 0,692 | 5,36 | 6,26 | 4,84 |
| 3422 | 0,872 | 0,794 | 0,700 | 0,676 | 0,793 | 8,53 | 8,85 | 8,16 |
| 3472 | 0,388 | 0,705 | 0,266 | 0,630 | 0,569 | 2,20 | 2,76 | 2,28 |
| 3492 | 0,465 | 0,411 | 0,600 | 0,722 | 0,600 | 3,00 | 2,99 | 0,75 |

Tableau 1 : Numéro d'ordre du fromage, performance par critère et note bruitée

D'un point de vue plus général, le nombre de critères est $m$ et il y a autant de poids tels que $1 \geq w_j \geq 0$ pour tout $j \epsilon \{1, \cdots m\}$, avec $\sum_{j=1}^{m} w_j = 1$. La théorie des ensembles flous a fourni des outils utiles pour représenter la préférence du décideur (ZADEH 1975) (GRABISH 1996), nous utiliserons le terme d'utilité $u_k(i)$ pour indiquer la valeur d'un critère $k$ entre 0 et 1 du produit $i$ comme donné tableau 1. Le score fournit par OWA par fromage s'écrit :

$$score_{OWA}(i) = \sum_{j=1}^{m} w_j u_{\sigma(j)}(i) \qquad (1)$$

Notons que les opérateurs OWA sont membres d'une famille plus générale d'opérateurs d'agrégation appelés intégrales de Sugeno et intégrales de Choquet (YAGER 1996) (YAGER 1998). Dans (RENAUD 2008) nous avons effectué la détermination des poids des performances, par identification paramétrique à partir de quatre échantillons classés, correspondant au mode de choix du décideur industriel. A présent nous allons faire apparaître trois aspects nouveaux : l'identification paramétrique à partir de notes données par le décideur et non plus par classement, valider la détermination des poids par identification paramétrique, puis comparer les résultats obtenus par identification paramétrique à partir de notes et de classements.

Malheureusement le décideur a estimé qu'il avait assez passé de temps lors du classement des 4 échantillons, et qu'il ne souhaitait pas donner de notes, ce qui lui aurait pris encore plus de temps (EINHORN 1974) (HAMMOND et al. 1975) (HAMMOND 2000). C'est pour cela que nous définirons un décideur « virtuel » dont les poids sont connus afin de vérifier si l'identification paramétrique permet de les retrouver. L'avantage de cette méthode est que nous allons pouvoir étudier l'intérêt de l'application d'outils issus des plans d'expériences, mais aussi d'autres modèles de décision comme l'intégrale de Choquet.

Afin de vérifier la présente approche, un décideur virtuel est proposé sur la base de l'opérateur OWA et de l'équation (1). Il attribue une note à chaque produit à l'aide des poids présentés dans le tableau 2. Pour un produit $i$ donné, la note est définie par l'expression suivante :

$$note(i) = 30 \, score_{OWA}(i) - 15 + e_k(i) \qquad (2)$$

Où $e_k$ est un aléa à distribution gaussienne centrée de variance égale à 0,1 (k=1), 0,5 (k=2) et 1 (k=3).

| Profil du décideur | Performance 1 | Performance 2 | Performance 3 | Performance 4 |
|---|---|---|---|---|
| Poids | 0,4 | 0,3 | 0,15 | 0,15 |

Tableau 2 : Ensemble des poids du décideur virtuel

Les scores obtenus pour chaque produit sont dans le tableau 1. L'utilisation de trois variances différentes permet d'en étudier l'influence. En effet dans le cas d'un véritable décideur, il existe plusieurs paramètres a priori inconnus comme ses préférences, la valeur à l'origine (ici -15) et l'échelle de notation qu'il utilise (dans notre cas un score de 0,9 correspond à une note de 12 sans aléa) ou l'erreur de notation (variance d'erreur). Les erreurs sont supposées indépendantes. Les notes ont été calculées pour chaque produit en utilisant les trois variances. Le même ensemble de pondérations est appliqué aux trois profils d'erreur due aux décideurs.

Le principe de l'identification paramétrique est de minimiser l'écart entre la note estimée par le modèle et la note donnée par le décideur pour l'ensemble des produits de l'échantillon. Le vecteur des notes données par le décideur est noté $\vec{S}$. Le problème est que l'équation (2) n'est pas connue dans le cas d'un décideur réel, nous ferons donc comme si nous ne la connaissions pas. Nous supposerons cependant que les notes données par le décideur sont du type, pour le fromage n°$i$ :

$$\left.\begin{array}{l}\overline{note}(i) = a\, score_{OWA}(i) + b = b + aw_1 u_{\sigma(1)}(i) + aw_2 u_{\sigma(2)}(i) + aw_3 u_{\sigma(3)}(i) + aw_4 u_{\sigma(4)}(i) \\ \overline{note}(i) = \begin{pmatrix} 1 & u_{\sigma(1)}(i) & u_{\sigma(2)}(i) & u_{\sigma(3)}(i) & u_{\sigma(4)}(i) \end{pmatrix} \begin{pmatrix} b \\ aw_1 \\ aw_2 \\ aw_3 \\ aw_4 \end{pmatrix} = \vec{x}^T(i)\vec{\theta} \end{array}\right\} \quad (3)$$

Où le vecteur des paramètres est $\vec{\theta} = \begin{pmatrix} b \\ aw_1 \\ aw_2 \\ aw_3 \\ aw_4 \end{pmatrix}$, ce qui permet de calculer les poids avec $\sum_{j=1}^{m} aw_j = a$ ($w_k = \frac{aw_k}{\sum_{j=1}^{m} aw_j}$), et que le vecteur des notes estimées est $\vec{\Sigma} = \begin{pmatrix} \overline{note}(1) \\ \vdots \\ \overline{note}(p) \end{pmatrix}$ si l'échantillon est formé de $p$ fromages. Nous avons donc, dans le cas de 4 critères :

$$\vec{\Sigma} = \begin{pmatrix} \vec{x}^T(1) \\ \vdots \\ \vec{x}^T(p) \end{pmatrix} \vec{\theta} = \bar{\bar{X}}\vec{\theta} = \begin{pmatrix} 1 & u_{\sigma(1)}(1) & u_{\sigma(2)}(1) & u_{\sigma(3)}(1) & u_{\sigma(4)}(1) \\ & & \vdots & & \\ 1 & u_{\sigma(1)}(m) & u_{\sigma(2)}(m) & u_{\sigma(3)}(m) & u_{\sigma(4)}(m) \end{pmatrix} \vec{\theta} \quad (4)$$

$\bar{\bar{X}}$ est la matrice des utilités classées par ordre de performance pour tous les produits de l'échantillon. L'ensemble optimal de poids est obtenu en minimisant, avec $\vec{S}^T = (note(1) \quad \cdots \quad note(p))$ :

$$\left.\begin{array}{l} E = (\vec{\Sigma} - \vec{S})^T(\vec{\Sigma} - \vec{S}) = (\bar{\bar{X}}\vec{\theta} - \vec{S})^T(\bar{\bar{X}}\vec{\theta} - \vec{S}) \\ E = \vec{\theta}^T \bar{\bar{X}}^T \bar{\bar{X}} \vec{\theta} - 2\vec{S}^T \bar{\bar{X}} \vec{\theta} + \vec{S}^T \vec{S} \\ \left(\frac{\partial E}{\partial \vec{\theta}}\right)^T = 2\bar{\bar{X}}^T \bar{\bar{X}} \vec{\theta} - 2\bar{\bar{X}}^T \vec{S} = 0 \end{array}\right\} \quad (5)$$

L'ensemble des poids estimés est donc :

$$\hat{\vec{\theta}} = (\bar{\bar{X}}^T \bar{\bar{X}})^{-1} \bar{\bar{X}}^T \vec{S} \quad (6)$$

Où $\bar{\bar{X}}^T \bar{\bar{X}}$ est la matrice d'information de Fischer. C'est cette matrice qui permet de définir le domaine de confiance des paramètres. De plus, $(\bar{\bar{X}}^T \bar{\bar{X}})^{-1} \bar{\bar{X}}^T$ est la pseudo inverse de $\bar{\bar{X}}$, mais si cette dernière est inversible alors $\hat{\vec{\theta}} = \bar{\bar{X}}^{-1} \vec{S}$. L'équation (6) est appliquée lorsqu'on utilise les notes données par le décideur virtuel présentées dans le tableau 1. Pour comparer les poids obtenus dans le cas où le décideur ne donne pas de note mais fournit un classement de plusieurs échantillons de fromage nous nous référerons à (RENAUD et al. 2008). C'est dans cette publication qu'est décrite la méthode permettant d'estimer les poids à partir d'un classement. Notons $u_{\sigma(\alpha)}^{\beta}$ l'utilité de la $\alpha^{ième}$ performance du fromage classé $\beta^{ième}$ dans l'échantillon, alors :

$$\begin{pmatrix} \hat{w}_1 \\ \hat{w}_2 \\ \hat{w}_3 \\ \hat{w}_4 \end{pmatrix} = \begin{pmatrix} u_{\sigma(1)}^1 - 2u_{\sigma(1)}^2 + u_{\sigma(1)}^3 & u_{\sigma(2)}^1 - 2u_{\sigma(2)}^2 + u_{\sigma(2)}^3 & u_{\sigma(3)}^1 - 2u_{\sigma(3)}^2 + u_{\sigma(3)}^3 & u_{\sigma(4)}^1 - 2u_{\sigma(4)}^2 + u_{\sigma(4)}^3 \\ u_{\sigma(1)}^2 - 2u_{\sigma(1)}^3 + u_{\sigma(1)}^4 & u_{\sigma(2)}^2 - 2u_{\sigma(2)}^3 + u_{\sigma(2)}^4 & u_{\sigma(3)}^2 - 2u_{\sigma(3)}^3 + u_{\sigma(3)}^4 & u_{\sigma(4)}^2 - 2u_{\sigma(4)}^3 + u_{\sigma(4)}^4 \\ u_{\sigma(1)}^3 - 2u_{\sigma(1)}^4 + u_{\sigma(1)}^5 & u_{\sigma(2)}^3 - 2u_{\sigma(2)}^4 + u_{\sigma(2)}^5 & u_{\sigma(3)}^3 - 2u_{\sigma(3)}^4 + u_{\sigma(3)}^5 & u_{\sigma(4)}^3 - 2u_{\sigma(4)}^4 + u_{\sigma(4)}^5 \\ 1 & 1 & 1 & 1 \end{pmatrix}^{-1} \begin{pmatrix} 0 \\ 0 \\ 0 \\ 1 \end{pmatrix} \quad (7)$$

Cette méthode donne directement la valeur des poids. Huit échantillons ont été sélectionnés pour cette publication, les quatre premiers étant tirés de (RENAUD et al. 2008).

| E1 | E2 | E3 | E4 | E5 | E6 | E7 | E8 |
|---|---|---|---|---|---|---|---|
| 613 | 1853 | 1853 | 1323 | 1853 | 2922 | 3062 | 813 |
| 2573 | 2663 | 613 | 703 | 592 | 2663 | 2343 | 1433 |
| 292 | 733 | 162 | 2902 | 2922 | 1793 | 2902 | 1923 |
| 162 | 2972 | 1103 | 1103 | 2663 | 1813 | 592 | 2483 |
| 3062 | 1103 | 3262 | 2083 | 162 | 3112 | 292 | 3472 |

Tableau 3 : Les 8 échantillons de test

Les échantillons E5 à E8 sont composés aléatoirement à partir de la même population de 47 produits. Il n'y a pas d'échantillons identiques. Nous allons donc extraire les poids des notes des fromages de ces 8 échantillons à partir des notes indiquées dans le tableau 1, et les comparer à ceux obtenus à partir des classements dans chaque échantillon.

La suite du présent article présente une méthode originale basée sur la D-optimalité pour déterminer un ensemble d'échantillons permettant de déterminer les paramètres de décision avec une meilleure précision. Cette méthode sera appliquée et validée dans une approche OWA en premier lieu.

Ensuite, nous remplacerons l'opérateur OWA par MAUT (Multi Attribute Utility Theory), en effet dans certains problèmes de décision multicritères il est difficile de choisir la méthode qui décrit le mieux le comportement du décideur. Nous pourrons donc comparer ces deux modèles de décision. Puis nous étudierons une approche par combinaison linéaire de OWA et MAUT. Ensuite l'approche par intégrale discrète de CHOQUET sera appliquée.

**4 Identification paramétrique des poids du modèle OWA à l'aide de notes**

Le calcul des huit ensembles de poids, pour les échantillons présentés tableau 3, est obtenu par identification paramétrique, à l'aide de l'équation (6), en fonction du profil du décideur. Le profil de décideur avec variance d'erreur égale à 0,1 est dans le tableau 4, celui avec une variance de 0,5 dans le tableau 5 et celui de variance 1 dans le tableau 6. Pour chaque ensemble de poids positifs, la moyenne quadratique des écarts entre les poids du décideur virtuel (tableau 2) et ceux des résultats d'identification paramétrique est calculé (racine de la moyenne des carrés des écarts).

| Poids | E1 | E2 | E3 | E4 | E5 | E6 | E7 | E8 |
|---|---|---|---|---|---|---|---|---|
| $W_1$ | 0,6016 | 0,0773 | -0,3278 | 0,3427 | 0,6302 | 0,3946 | 0,9264 | 0,7256 |
| $W_2$ | 0,0617 | 0,4837 | -0,3782 | 0,2642 | 0,3287 | 0,2428 | -0,2020 | -0,2778 |
| $W_3$ | 0,1737 | -0,1434 | 1,3593 | 0,2609 | 0,3460 | -0,0979 | 0,4453 | 0,2952 |
| $W_4$ | 0,1630 | 0,5823 | 0,3467 | 0,1322 | -0,3049 | 0,4605 | -0,1697 | 0,2570 |
| Moy. Quad. | 0,1566 | | | 0,0655 | | | | |

Tableau 4 : Poids obtenus avec les notes du décideur virtuel avec variance égale à 0,1

| Poids | E1 | E2 | E3 | E4 | E5 | E6 | E7 | E8 |
|---|---|---|---|---|---|---|---|---|
| $W_1$ | 0,5936 | 0,5003 | 2,0802 | 0,7690 | 0,5760 | 0,5638 | -2,6470 | 0,1775 |
| $W_2$ | 0,0221 | 0,1532 | 2,6998 | -0,3979 | 0,2442 | -0,0378 | 3,3176 | 0,7378 |
| $W_3$ | 0,2529 | 0,2811 | -3,4976 | 0,8055 | 0,2681 | -0,3926 | -1,0076 | 0,0709 |
| $W_4$ | 0,1314 | 0,0654 | -0,2824 | -0,1765 | -0,0883 | 0,8666 | 1,3371 | 0,0137 |
| Moy. Quad. | 0,1772 | 0,1183 | | | | | | 0,2487 |

Tableau 5 : Poids obtenus avec les notes du décideur virtuel avec variance égale à 0,5

Nous constatons que les poids obtenus avec les échantillons E3, E5, E6 et E7 ont toujours des valeurs négatives, or cela n'a aucun sens d'accepter des valeurs négatives de certains poids. C'est pourquoi la moyenne quadratique n'est pas calculée pour les échantillons présentant un ou plusieurs poids négatifs. Seuls les échantillons donnant des valeurs positives pour tous les poids présentent un intérêt. Ainsi, la façon de sélectionner les produits pour former un échantillon destiné à l'identification paramétrique est très important.

La moyenne quadratique est la racine carrée de la moyenne des carrés des écarts entre les poids du décideur virtuel (tableau 2) et ceux identifiés dans les tableaux 4, 5 et 6.

| Poids | E1 | E2 | E3 | E4 | E5 | E6 | E7 | E8 |
|---|---|---|---|---|---|---|---|---|
| $W_1$ | 0,2210 | -0,0810 | 1,0887 | -0,4752 | 0,3641 | 0,4486 | 2,1961 | 0,9519 |
| $W_2$ | 0,4160 | 0,5600 | 1,9583 | 2,6811 | -0,1288 | 0,1413 | -2,6155 | -0,2129 |
| $W_3$ | 0,0902 | 0,0557 | -2,0497 | -2,3537 | 0,0872 | -0,1456 | -0,2354 | 0,5813 |
| $W_4$ | 0,2728 | 0,4653 | 0,0028 | 1,1478 | 0,6775 | 0,5557 | 1,6548 | -0,3203 |
| Moy. Quad. | 0,1266 | | | | | | | |

Tableau 6 : Poids obtenus avec les notes du décideur virtuel avec variance égale à 1

Plus la moyenne quadratique est faible plus les poids estimés sont proches de ceux du décideur virtuel. L'idéal est de choisir l'échantillon donnant la moyenne quadratique la plus faible, mais pour cela il faut avoir effectué l'identification paramétrique. Cette moyenne quadratique est la plus faible pour l'échantillon E4 avec les notes du décideur virtuel avec variance égale à 0,1. Cependant, cet échantillon ne donne pas de bons résultats avec les notes ayant une variance 0,5 ou 1. L'échantillon donnant des poids positifs dans tous les cas est E1, avec une moyenne quadratique proche de 0,15 à 0,03 près, ce qui est honorable. Les échantillons E2, E4 et E8 ne donnent des poids positifs que pour une seule variance, pas toujours la plus faible. Rappelons que plus la variance est faible plus la note attribuée à un produit est peu entachée d'erreur. Le seul effet significatif de la variance est que pour la valeur 1 un seul échantillon sur 8 peut donner des valeurs acceptables des poids.

**5 Identification paramétrique des poids du modèle OWA à l'aide de classements**

Dans (RENAUD et al. 2008) l'identification des poids était faite à partir de classements faits par le décideur vrai. Ici, nous ne pouvons pas demander cela au vrai décideur et avons décidé d'utiliser un décideur virtuel afin d'apprécier l'intérêt des méthodes appliquées. C'est donc à partir des notes calculées avec des variances d'erreur allant de 0,1 à 1 que nous classons les fromages d'un échantillon (tableau 1). Cependant, l'erreur ajoutée à la note théorique fait que le classement d'un échantillon peut différer selon la valeur de la variance, et donc que les poids calculés sont différents.

| Poids | E1 | E2 | E3 | E4 | E5 | E6 | E7 | E8 |
|---|---|---|---|---|---|---|---|---|
| $W_1$ | 0,4914 | 0,5253 | -1,6746 | 0,4249 | 0,5761 | 0,3710 | 1,7655 | -1,6383 |
| $W_2$ | 0,0718 | 0,2197 | -1,8904 | 0,1417 | 0,3278 | 0,2635 | -0,9771 | 4,0681 |
| $W_3$ | 0,2313 | 0,2357 | 3,9386 | 0,3602 | 0,2785 | -0,1421 | 0,9061 | 0,5168 |
| $W_4$ | 0,2054 | 0,0193 | 0,6265 | 0,0732 | -0,1823 | 0,5077 | -0,6945 | -1,9466 |
| Moy. Quad. | 0,1324 | 0,1079 | | 0,1376 | | | | |

Tableau 7 : Poids obtenus avec le classement du décideur virtuel avec variance égale à 0,1

| Poids | E1 | E2 | E3 | E4 | E5 | E6 | E7 | E8 |
|---|---|---|---|---|---|---|---|---|
| $W_1$ | 0,4914 | 0,5253 | 6,8544 | 0,4249 | 0,6358 | 0,2335 | -5,6808 | 0,1972 |
| $W_2$ | 0,0718 | 0,2197 | 9,0143 | 0,1417 | 0,1274 | 0,4351 | 6,3862 | 1,0534 |
| $W_3$ | 0,2313 | 0,2357 | -13,2177 | 0,3602 | 0,3032 | -0,1072 | -2,3740 | 0,3652 |
| $W_4$ | 0,2054 | 0,0193 | -1,6510 | 0,0732 | -0,0663 | 0,4386 | 2,6686 | -0,6158 |
| Moy. Quad. | 0,1324 | 0,1079 | | 0,1376 | | | | |

Tableau 8 : Poids obtenus avec le classement du décideur virtuel avec variance égale à 0,5

Là encore, les moyennes quadratiques ne sont calculées que si la totalité des poids estimés sont positifs. Or seuls les échantillons E1, E2 et E4 fournissent des poids acceptables dans le cas de variance inférieure à 1, et E1 quel que soit la variance d'erreur.

| Poids | E1 | E2 | E3 | E4 | E5 | E6 | E7 | E8 |
|---|---|---|---|---|---|---|---|---|
| $W_1$ | 0,0215 | -0,3319 | 6,8544 | -0,7436 | 0,5761 | 0,4345 | 2,5685 | 0,1972 |
| $W_2$ | 0,6337 | 0,8303 | 9,0143 | 1,3392 | 0,3278 | 0,1562 | -2,1092 | 1,0534 |
| $W_3$ | 0,1401 | -0,1214 | -13,2177 | -0,7440 | 0,2785 | -0,1666 | 1,2524 | 0,3652 |
| $W_4$ | 0,2046 | 0,6231 | -1,6510 | 1,1484 | -0,1823 | 0,5759 | -0,7117 | -0,6158 |
| Moy. Quad. | 0,2538 | | | | | | | |

Tableau 9 : Poids obtenus avec le classement du décideur virtuel avec variance égale à 1

Globalement, en comparant l'identification paramétrique à partir de notes et de classements il est difficile de dire qu'une méthode est meilleure que l'autre, les résultats étant assez semblables. Nous verrons si l'apport de l'amélioration des échantillons par D-optimalité y change quelque chose.

**6 Amélioration de l'identification paramétrique à l'aide de la D-optimalité**

Nous avons vu en sections 4 et 5 que tous les échantillons ne permettent pas une identification paramétrique satisfaisante. Ainsi, notre objectif est de proposer une méthode de sélection des produits les plus représentatifs au sein de l'ensemble des produits. Nous avons rappelé, section 3, que la matrice d'information de Fischer, $\bar{\bar{X}}^T\bar{\bar{X}}$, permet de définir le domaine de confiance des paramètres. Afin de réduire la taille du domaine de confiance, c'est-à-dire l'incertitude sur la valeur des paramètres, il faut diminuer le déterminant de $(\bar{\bar{X}}^T\bar{\bar{X}})^{-1}$, donc qu'il faut maximiser la valeur du déterminant de $\bar{\bar{X}}^T\bar{\bar{X}}$. C'est ce qu'on appelle la D-optimalité, la sélection des produits de l'échantillon à privilégier pouvant se faire grâce aux algorithmes DETMAX ou Fedorov (WALTER et al. 1994). Ainsi, le principe de la méthode est de minimiser l'intervalle de confiance des paramètres. Une forme itérative simple est donnée par :

1) Choisir arbitrairement $p$ produits pour former la 1ère matrice $\bar{\bar{X}}$.
2) Chercher hors de $\bar{\bar{X}}$ le vecteur $\vec{y}$ qui maximise $u_1 = \vec{y}^T(\bar{\bar{X}}^T\bar{\bar{X}})^{-1}\vec{y}$, puis dans $\bar{\bar{X}}$ le vecteur $\vec{z}$ qui minimise $u_2 = \vec{z}^T(\bar{\bar{X}}^T\bar{\bar{X}})^{-1}\vec{z} - \frac{(\vec{y}^T(\bar{\bar{X}}^T\bar{\bar{X}})^{-1}\vec{z})^2}{1+u_1}$.
3) Si $(1+u_1)(1-u_2) \leq 1$ la procédure s'arrête (la matrice $\bar{\bar{X}}$ est optimale car il n'est plus possible d'augmenter le déterminant de $\bar{\bar{X}}^T\bar{\bar{X}}$), sinon on substitue le vecteur $\vec{y}$ au vecteur $\vec{z}$ dans $\bar{\bar{X}}$ pour obtenir la nouvelle matrice $\bar{\bar{X}}$ et inversement hors de $\bar{\bar{X}}$, puis retour à 2).

Les 8 échantillons du tableau 3 sont les points de départ de l'algorithme et le tableau 10 donne les échantillons obtenus à l'aide des performances de chaque produit (OWA). L'objectif est d'obtenir l'échantillon ayant la matrice $\bar{\bar{X}}^T\bar{\bar{X}}$ dont le déterminant est le plus grand possible. Dans le tableau 10, nous constatons que les échantillons résultats E'1, E'2, E'4, E'5, E'6, E'7 et E'8 sont tous identiques, soit 7 sur 8, et donnent le déterminant de $\bar{\bar{X}}^T\bar{\bar{X}}$ le plus élevé, soit 0,01841721.

| Echantillon initial | E1 | E2 | E3 | E4 | E5 | E6 | E7 | E8 |
|---|---|---|---|---|---|---|---|---|
| Déterminant initial | 1,75E-03 | 3,52E-04 | 2,16E-06 | 1,29E-05 | 1,32E-03 | 1,69E-04 | 3,27E-05 | 9,51E-07 |
| Nombre d'itérations | 4 | 5 | 3 | 6 | 3 | 4 | 4 | 5 |
| Déterminant final | 1,84E-02 | 1,84E-02 | 1,54E-02 | 1,84E-02 | 1,84E-02 | 1,84E-02 | 1,84E-02 | 1,84E-02 |
| Echantillon final | E'1 | E'2 | E'3 | E'4 | E'5 | E'6 | E'7 | E'8 |
| Composé | 592 | 292 | 813 | 813 | 292 | 3162 | 162 | 813 |
|  | 813 | 592 | 613 | 292 | 592 | 813 | 813 | 162 |
|  | 292 | 813 | 162 | 592 | 3162 | 592 | 3162 | 3162 |
|  | 162 | 3162 | 3162 | 162 | 813 | 292 | 592 | 592 |
|  | 3162 | 162 | 292 | 3162 | 162 | 162 | 292 | 292 |

Tableau 10 : Echantillons améliorés par D-optimalité pour OWA

L'échantillon obtenu, noté E', est formé des produits 162, 292, 592, 813 et 3162. C'est cet échantillon qui doit être testé par identification paramétrique, à l'aide des notes présentées tableau 1, d'une part et le classement correspondant d'autre part. Nous constatons que le classement est identique pour les notes avec les trois variances d'erreur, donc les poids obtenus ne dépendent pas de cette variance.

| OWA | Variance 0.1 | Variance 0.5 | Variance 1 | Classement |
|---|---|---|---|---|
| Poids 1 | 0,53957671 | 0,4665228 | 0,4418882 | 0,67176409 |
| Poids 2 | 0,20303979 | 0,2150367 | 0,275756 | 0,09072304 |
| Poids 3 | 0,1809982 | 0,21212897 | 0,1337481 | 0,22273785 |
| Poids 4 | 0,0763853 | 0,10631154 | 0,14860771 | 0,01477502 |
| Moy. Quad. | 0,09389216 | 0,06597866 | 0,02553652 | 0,18790234 |

Tableau 11 : Poids obtenus avec E' à partir des notes et du classement

Nous constatons que les poids obtenus à partir du classement sont entachés de plus d'erreur que ceux obtenus à partir des notes. Il n'y a pas d'amélioration par D-optimalité, bien au contraire. Les poids obtenus à partir des notes sont plus proches de ceux du décideur virtuel (tableau 2) que ceux présentés dans les tableaux 4, 5, et 6 (moyennes quadratiques d'écart nettement plus faibles). Curieusement, les poids obtenus s'améliorent un peu quand la variance d'erreur sur les notes augmente, ce phénomène demanderait à être vérifié. Cependant, il est indéniable que l'amélioration des échantillons par D-optimalité donne de bien meilleurs résultats en ce qui concerne l'identification à partir des notes. De plus, l'équation (3) donne un modèle d'appréciation des notes par le décideur, avec :

| OWA | Variance 0.1 | Variance 0.5 | Variance 1 |
|---|---|---|---|
| $b$ | -26,5053606 | -17,0844142 | -19,0984469 |
| $a$ | 43,0501528 | 31,5819536 | 34,893703 |

Tableau 12 : modèle d'établissement des notes

Les valeurs de $b = -15$ et $a = 30$ sont données équation (2) pour le décideur virtuel. Nous ne retrouvons pas parfaitement ces valeurs dans le tableau 12, mais juste dans les grandes lignes. C'est dans le cadre de la variance d'erreur 0,5 que nous les approchons le plus.

Nous avons créé un décideur virtuel à l'aide de la méthode OWA, équations (1) et (2), et avons ensuite identifié les poids d'un modèle lui aussi basé sur OWA, équation (3). Nous avons donc fait l'hypothèse que nous connaissions le modèle représentant l'exacte réalité de la façon de décider du professionnel. Pourtant dans la réalité c'est très rarement le cas, nous devons donc vérifier ce qui se passe si nous nous trompons de modèle (SHANTEAU 1995) (SHANTEAU et al. 2002). Pour cela nous allons voir ce qui se passe si nous choisissons un modèle de type MAUT (Multi Attribute Utility Theory). La différence entre les deux modèles est que les poids s'appliquent à des valeurs de critère dans MAUT alors qu'ils s'appliquent à une performance dans OWA. Dans le cas de OWA la meilleure performance correspond au critère ayant la valeur normalisée (CHOQUET 1953) la plus élevée, quel que soit ce critère, et inversement pour la moins bonne performance. Ainsi, le score selon OWA est donné par l'équation (1) et celui de MAUT par :

$$score_{MAUT}(i) = \sum_{j=1}^{m} w_j u_j(i) \qquad (8)$$

Où $i$ représente le produit à évaluer, $j$ le critère considéré et son poids associé $w_j$, et $u_j(i)$ l'utilité correspondante du critère $j$ du produit $i$, c'est-à-dire la valeur normalisée entre 0 et 1 de la valeur du critère.

L'objectif de la théorie de l'utilité multi-attributs (MAUT) est de modéliser les préférences du décideur, représentées au moyen d'une fonction d'utilité globale $u$. L'approche d'amélioration par D-optimalité est également appliquée à l'ensemble des huit échantillons, de E1 à E8. Cependant, avant cela il convient d'indiquer les résultats de l'application de MAUT aux 8 échantillons précités. Le tableau 13 précise ces résultats. Nous constatons qu'aucun des 8 échantillons initiaux donne des poids positifs (donc acceptables) pour les deux types d'identification paramétrique (à partir des notes et à partir des classements) et pour toutes les variances d'erreur de notation (les classements étant définis à partir des notes). Pire, les échantillons E5 et E8 ne fournissent aucun jeu de poids acceptable (0 sur 6). Seul l'échantillon E4 n'a qu'un jeu de poids inacceptable sur six. L'échantillons E6 n'a qu'un seul jeu de poids acceptable sur 6, obtenu par identification paramétrique à partir des classements. L'échantillon E7 a 2 jeux de poids acceptables sur 6, obtenus l'un à partir de notes et l'autre à partir d'un classement. Notons que ces deux jeux de poids sont très différents. Les échantillons E1, E2 et E3 ont 4 jeux de poids acceptables sur 6, cependant à valeurs très différentes pour E1.

Les résultats de l'application de la D-optimalité sont présentés dans le tableau 14. L'ensemble des échantillons améliorés va de E"1 à E"8. L'échantillon E"4 a la valeur de déterminant la plus élevée, et c'est donc l'échantillon sélectionné pour calculer les poids par identification paramétrique dans le cas du décideur virtuel. Or E''4 est obtenu à partir de E4, celui reconnu ci-dessus comme le meilleur échantillon initial (1 jeu de poids inacceptable sur 6).

| MAUT | | Identification par notes | | | Identification par classement | | |
|---|---|---|---|---|---|---|---|
| | | V=0.1 | V=0.5 | V=1 | V=0.1 | V=0.5 | V=1 |
| E1 | b | -4,1717 | -5,5688 | -8,7936 | | | |
| | a | 17,547 | 19,262 | 23,789 | | | |
| | $W_1$ | -0,0781 | -0,1364 | 0,3802 | 0,0391 | 0,0391 | 0,5018 |
| | $W_2$ | 0,5460 | 0,5331 | 0,2412 | 0,4135 | 0,4135 | 0,1344 |
| | $W_3$ | 0,1908 | 0,3092 | 0,0884 | 0,2471 | 0,2471 | 0,1625 |
| | $W_4$ | 0,3413 | 0,2941 | 0,2901 | 0,3003 | 0,3003 | 0,2012 |
| E2 | b | -7,8651 | -19,728 | 0,1889 | | | |
| | a | 21,951 | 41,670 | 9,6562 | | | |
| | $W_1$ | 0,3191 | 0,2674 | 0,1870 | 0,2700 | 0,2700 | 0,3523 |
| | $W_2$ | 0,2625 | 0,1371 | 0,7878 | 0,0828 | 0,0828 | -0,5649 |
| | $W_3$ | 0,3234 | 0,2178 | 0,7524 | 0,2697 | 0,2697 | -0,6211 |
| | $W_4$ | 0,0950 | 0,3777 | -0,7272 | 0,3776 | 0,3776 | 1,8337 |
| E3 | b | -9,910 | -28,585 | -22,974 | | | |
| | a | 25,633 | 53,082 | 44,230 | | | |
| | $W_1$ | 0,0524 | 0,4169 | 0,4604 | 0,5261 | -9,684 | -9,684 |
| | $W_2$ | 0,4051 | 0,2391 | 0,1946 | 0,2259 | 6,918 | 6,918 |
| | $W_3$ | 0,2888 | 0,1599 | 0,1417 | 0,1042 | 2,684 | 2,684 |
| | $W_4$ | 0,2537 | 0,1841 | 0,2033 | 0,1438 | 1,081 | 1,081 |
| E4 | b | -12,75 | -8,022 | -6,726 | | | |
| | a | 29,52 | 21,84 | 23,37 | | | |
| | $W_1$ | 0,2211 | 0,1981 | -0,1543 | 0,2160 | 0,2160 | 0,3056 |
| | $W_2$ | 0,2437 | 0,1234 | 0,4454 | 0,2277 | 0,2277 | 0,3471 |
| | $W_3$ | 0,2234 | 0,2760 | 0,4349 | 0,2312 | 0,2312 | 0,2661 |
| | $W_4$ | 0,3118 | 0,4025 | 0,2739 | 0,3251 | 0,3251 | 0,0812 |
| E5 | b | 9,245 | 10,437 | 9,6227 | | | |
| | a | 3,511 | 1,5331 | 3,095 | | | |
| | $W_1$ | -0,9288 | -5,171 | -1,765 | 0,8899 | 0,8621 | 0,8899 |
| | $W_2$ | 0,2683 | -0,7134 | 0,3480 | -0,0210 | 0,1363 | -0,0210 |
| | $W_3$ | -0,9223 | 1,626 | -0,7933 | 1,0085 | 0,4627 | 1,0085 |
| | $W_4$ | 2,5827 | 5,258 | 3,210 | -0,8774 | -0,4611 | -0,8774 |
| E6 | b | -51,54 | -3,174 | -60,18 | | | |
| | a | 88,60 | 13,67 | 103,45 | | | |
| | $W_1$ | 0,5183 | -0,0558 | 0,5410 | 0,5327 | 0,4800 | 0,5543 |
| | $W_2$ | 0,2116 | 0,0077 | 0,2345 | 0,2169 | 0,2066 | 0,2409 |
| | $W_3$ | -0,0921 | 1,0318 | -0,1641 | -0,1061 | 0,0250 | -0,1836 |
| | $W_4$ | 0,3622 | 0,0163 | 0,3886 | 0,3565 | 0,2884 | 0,3885 |
| E7 | b | -9,484 | -10,56 | -6,242 | | | |
| | a | 22,61 | 23,98 | 22,44 | | | |
| | $W_1$ | 0,6755 | 0,6405 | -0,1706 | 1,0776 | 0,7384 | 0,0844 |
| | $W_2$ | 0,1182 | 0,1199 | 0,5204 | -0,1036 | 0,1108 | 0,2977 |
| | $W_3$ | 0,1803 | -0,0864 | 0,2933 | 0,3311 | -0,2929 | 0,4757 |
| | $W_4$ | 0,0259 | 0,3261 | 0,3569 | -0,3051 | 0,4437 | 0,1422 |

Tableau 13a : Résultat obtenus avec MAUT sur 7 échantillons initiaux

| MAUT | | Identification par notes | | | Identification par classement | | |
|---|---|---|---|---|---|---|---|
| | | V=0,1 | V=0,5 | V=1 | V=0,1 | V=0,5 | V=1 |
| E8 | b | -7,036 | -3,452 | 3,089 | | | |
| | a | 28,93 | 30,21 | 36,88 | | | |
| | $W_1$ | 1,772 | 2,803 | 4,417 | -4,152 | 6,368 | 6,368 |
| | $W_2$ | 0,9559 | 1,535 | 2,228 | -2,348 | 3,303 | 3,303 |
| | $W_3$ | -0,1275 | -0,3816 | -0,6867 | 0,849 | -1,109 | -1,109 |
| | $W_4$ | -1,600 | -2,957 | -4,958 | 6,651 | -7,562 | -7,562 |

Tableau 13b : Résultats avec MAUT pour l'échantillon E8

Il est normal que les résultats diffèrent entre OWA et MAUT car la matrice $\bar{\bar{X}}$ n'est pas la même dans les deux cas, y compris pour les échantillons initiaux puisque l'ordre dans lequel les valeurs normalisées des critères sont écrites est différent (ordre des performances et ordre des critères).

| Echantillon initial | E1 | E2 | E3 | **E4** | E5 | E6 | E7 | E8 |
|---|---|---|---|---|---|---|---|---|
| Déterminant initial | 1,12E-03 | 2,64E-04 | 1,19E-04 | **2,03E-04** | 1,06E-05 | 3,55E-06 | 8,11E-05 | 1,06E-08 |
| Nombre d'itérations | 4 | 3 | 3 | **3** | 3 | 4 | 3 | 4 |
| Déterminant final | 2,43E-02 | 2,66E-02 | 1,32E-02 | **3,10E-02** | 2,22E-02 | 2,97E-02 | 2,71E-02 | 2,21E-02 |
| Echantillon final | E''1 | E''2 | E''3 | **E''4** | E''5 | E''6 | E''7 | E''8 |
| Composition des échantillons | 1193 | 2343 | 3162 | 1323 | 733 | 3162 | 3062 | 813 |
| | 733 | 3062 | 613 | 733 | 592 | 2663 | 162 | 3062 |
| | 1323 | 733 | 162 | 162 | 1103 | 1193 | 733 | 162 |
| | 162 | 162 | 1103 | 1103 | 2663 | 733 | 592 | 733 |
| | 3062 | 1103 | 2663 | 3062 | 162 | 162 | 1103 | 3472 |

Tableau 14 : Echantillons améliorés par D-optimalité pour MAUT

| MAUT | Variance 0,1 | Variance 0,5 | Variance 1 | Classement |
|---|---|---|---|---|
| Poids 1 | 0,29996248 | 0,19295036 | 0,24667935 | -0,85647455 |
| Poids 2 | 0,23774973 | 0,30137857 | 0,24422595 | 2,28906322 |
| Poids 3 | 0,24357378 | 0,28100546 | 0,24588653 | 0,42772799 |
| Poids 4 | 0,21871401 | 0,2246656 | 0,26320817 | -0,86031665 |

Tableau 15 : Poids obtenus avec E''4 à partir des notes et du classement

Les résultats de l'identification paramétrique dans le cas de MAUT, à partir des notes dans le cas des trois variances et à partir du classement, sont donnés tableau 15. Là encore, le classement des produits à l'aide des notes avec les trois variances d'erreur est identique, et donc indépendant de cette variance d'erreur des notes. Nous constatons que les poids obtenus par le classement ne sont pas valides puisque certains ont des valeurs négatives. Les poids obtenus à l'aide des notes (tableau 1) sont cohérents entre eux et pratiquement toujours compris entre 0,2 et 0,3. Cependant, ils sont très différents de ceux du décideur virtuel tableau 2. Ceci est normal puisque le décideur virtuel est basé sur OWA alors que le modèle utilisé pour identifier les poids est MAUT. C'est pourquoi nous ne calculons pas ici la moyenne quadratique des écarts entre les poids du décideur virtuel et ceux identifiés. Pour juger de l'efficacité de l'identification paramétrique nous comparerons le score obtenu par MAUT pour les produits de E''4 avec les notes données tableau 1 pour les trois variances d'erreur. Là encore le modèle d'attribution des notes par le décideur est donné par :

$$\left. \begin{array}{l} \overline{note}(i) = a\, score_{MAUT}(i) + b = b + aw_1 u_1(i) + aw_2 u_2(i) + aw_3 u_3(i) + aw_4 u_4(i) \\ \overline{note}(i) = \begin{pmatrix} 1 & u_1(i) & u_2(i) & u_3(i) & u_4(i) \end{pmatrix} \begin{pmatrix} b \\ aw_1 \\ aw_2 \\ aw_3 \\ aw_4 \end{pmatrix} = \vec{x}^T(i)\vec{\theta} \end{array} \right\} \quad (9)$$

Et l'identification paramétrique se fait toujours par l'équation (6). Nous obtenons donc également le tableau 16. Les résultats montrent que les coefficients obtenus restent relativement proches de ceux du décideur virtuel, malgré la différence existante entre OWA et MAUT. C'est dans le cas d'une variance d'erreur égale à 1 que l'écart est le plus grand avec le décideur virtuel. Ainsi MAUT permet de modéliser le comportement décisionnel du décideur virtuel aussi bien que OWA, mais avec des poids très différents. De plus, le tableau 15 montre que la relation entre les notes et le score MAUT est proche de l'équation (2) avec simplement des coefficients un petit peu plus faibles en valeur absolue. Nous pouvons donc dire que MAUT modélise bien la façon de noter du décideur (virtuel), tout comme OWA, mais avec des paramètres a et b bien plus proches de ceux du décideur virtuel.

| MAUT | Variance 0.1 | Variance 0.5 | Variance 1 |
|---|---|---|---|
| $b$ | -11,7312996 | -11,6403299 | -9,2451311 |
| $a$ | 27,7680234 | 27,9338988 | 23,9676348 |

Tableau 16 : modèle d'établissement des notes

### 7 Modélisation de la façon de noter du décideur par combinaison linéaire OWA-MAUT

Pour mieux représenter le mode de notation d'un décideur réel, nous pouvons être tenté d'utiliser comme modèle une combinaison linéaire de OWA et de MAUT. Nous allons voir dans la suite ce que donne cette méthode sur le décideur virtuel défini en section 3. Dans un premier temps, la méthode que nous proposerons est basé sur l'identification paramétrique du modèle :

$$\overline{note}(i) = a_1\, score_{OWA}(i) + a_2\, score_{MAUT}(i) + b =$$
$$b + a_1 w_1^1 u_{\sigma(1)}(i) + a_1 w_2^1 u_{\sigma(2)}(i) + a_1 w_3^1 u_{\sigma(3)}(i) + a_1 w_4^1 u_{\sigma(4)}(i) + a_2 w_1^2 u_1(i) + a_2 w_2^2 u_2(i) + a_2 w_3^2 u_3(i) + a_2 w_4^2 u_4(i)$$

$$\overline{note}(i) = \begin{pmatrix} 1 & u_{\sigma(1)}(i) & u_{\sigma(2)}(i) & u_{\sigma(3)}(i) & u_{\sigma(4)}(i) & u_1(i) & u_2(i) & u_3(i) & u_4(i) \end{pmatrix} \begin{pmatrix} b \\ a_1 w_1^1 \\ a_1 w_2^1 \\ a_1 w_3^1 \\ a_1 w_4^1 \\ a_2 w_1^2 \\ a_2 w_2^2 \\ a_2 w_3^2 \\ a_2 w_4^2 \end{pmatrix} = \vec{x}^T(i)\vec{\theta} \quad (10)$$

Où $w_\alpha^\beta$ est le poids d'indice $\alpha$ (performance $\alpha$ pour OWA et critère $\alpha$ pour MAUT) et $\beta = 1$ correspond à OWA alors que $\beta = 2$ correspond à MAUT. L'identification paramétrique est toujours réalisée à l'aide de l'équation (6). Il s'agit donc d'un modèle à 9 paramètres, nécessitant un échantillon de 9 fromages au minimum. Le modèle OWA a été identifié à l'aide de l'ensemble E' et MAUT par E''4, donc nous réunissons ces deux échantillons en un seul comportant 5+5-1=9 fromages car le produit 162 est commun aux deux premiers échantillons et nous ne devons pas avoir deux fois le même fromage dans l'ensemble d'identification. Le problème est que la matrice $\bar{\bar{X}} = \begin{pmatrix} \vec{x}^T(1) \\ \vdots \\ \vec{x}^T(9) \end{pmatrix}$ a un déterminant égal à $2{,}34\, 10^{-20}$, autant dire qu'elle est non inversible. Forcer l'inversion conduit à des paramètres absurdes. Il faut dire que dans chaque ligne on trouve deux fois la même utilité, bien que pas dans le même ordre. Nous avons donc étendu l'ensemble d'identification à 28 fromages sur 47 sans plus de succès car le déterminant de $\bar{\bar{X}}^T \bar{\bar{X}}$ est alors de $-1{,}157\, 10^{-15}$ (choix des 28 produits précisé plus loin). C'est pourquoi nous repartirons des tableaux 11, 12, 14 et 15, c'est-à-dire des résultats de la section précédente, et nous poserons :

$$\left. \begin{array}{l} \overline{note}(i) = \omega \overline{note}_{OWA}(i) + (1-\omega)\overline{note}_{MAUT} \\ \overline{note}_{OWA} = a_1 score_{OWA}(i) + b_1 \\ \overline{note}_{MAUT} = a_2 score_{MAUT}(i) + b_2 \end{array} \right\} \quad (11)$$

Où $a_1$ et $b_1$ sont donnés tableau 12 et les poids pour le calcul des scores tableau 11, et où $a_2$ et $b_2$ sont donnés tableau 16 et les poids pour le calcul des scores tableau 15.

Nous effectuons l'optimisation de $\omega$ à l'aide des 9 fromages indiqués au début de cette section, c'est-à-dire ceux de E' et de E''4, sans doublon, grâce à :

$$\left. \begin{array}{l} E = \sum_{i=1}^{9}\left(\omega \overline{note}_{OWA}(i) + (1-\omega)\overline{note}_{MAUT}(i) - note(i)\right)^2 \\ \frac{dE}{d\omega} = 0 = 2\omega \sum_{i=1}^{9}\left(\overline{note}_{OWA}(i) - \overline{note}_{MAUT}(i)\right)^2 \\ -2\sum_{i=1}^{9}\left(\overline{note}_{OWA}(i) - \overline{note}_{MAUT}(i)\right)\left(note(i) - \overline{note}_{MAUT}(i)\right) \end{array} \right\} \quad (12a)$$

$$\omega = \frac{\sum_{i=1}^{9}\left(\overline{note}_{OWA}(i) - \overline{note}_{MAUT}(i)\right)\left(note(i) - \overline{note}_{MAUT}(i)\right)}{\sum_{i=1}^{9}\left(\overline{note}_{OWA}(i) - \overline{note}_{MAUT}(i)\right)^2} \quad (12b)$$

Les résultats sont présentés tableaux 17 et 18. Certaines notes simulées sont rigoureusement égales aux notes du décideur virtuel (tableau 1), ceci est dû à l'identification paramétrique avec une matrice $\bar{\bar{X}}$ inversible. Ce sont ces résultats qui sont utilisés pour obtenir $\omega$.

| Produit | OWA v=0.1 | OWA v=0.5 | OWA v=1 | MAUT 0.1 | MAUT 0.5 | MAUT 1 |
|---|---|---|---|---|---|---|
| 813 | 9,52 | 8,91 | 9,67 | 9,95818406 | 9,88157556 | 9,5539571 |
| 162 | 5,12 | 5,23 | 4,48 | 5,12 | 5,23 | 4,48 |
| 3162 | 2,77 | 3 | 2,64 | 1,8112674 | 1,83895927 | 3,00040151 |
| 592 | 10,5 | 9,64 | 10,15 | 10,7732418 | 10,5787608 | 9,88020608 |
| 292 | 7,37 | 6,55 | 7,31 | 6,21077388 | 6,17417799 | 5,73500338 |
| 733 | 5,58551697 | 5,80305472 | 5,60260059 | 5,2 | 6,92 | 6,09 |
| 1103 | 1,35176146 | 2,77452828 | 2,35356498 | 3,77 | 3,43 | 4,01 |
| 1323 | 11,0231588 | 9,93946218 | 10,4292886 | 10,52 | 9,98 | 9,77 |
| 3062 | 0,76959592 | 2,30338483 | 1,82365663 | 3,11 | 2,71 | 3,61 |

Tableau 17 : Notes simulées par MAUT et OWA pour les 3 variances d'erreur

|  | Variance 0.1 | Variance 0.5 | Variance 1 |
|---|---|---|---|
| $\omega$ | 0,1774339 | 0,64250741 | 0,28985627 |
| $1 - \omega$ | 0,8225661 | 0,35749259 | 0,71014373 |

Tableau 18 : Valeur de $\omega$ et $1 - \omega$ pour les 3 variances d'erreur

Nous constatons que pour les variances d'erreur 0.1 et 1 la valeur de $\omega$, c'est-à-dire le facteur associé à OWA est inférieur à 0,3 alors que le décideur virtuel a été basé sur OWA. Ce résultat surprenant est probablement dû aux erreurs ajoutées volontairement aux notes du décideur virtuel pour faire plus « vrai ». Il reste à vérifier la validité de cette combinaison linéaire par comparaison avec OWA et MAUT. Nous ferons la comparaison en calculant la variance d'écart entre les notes données par le décideur virtuel et celle simulées, ainsi que l'écart type correspondant. Cette comparaison sera faite sur les 28 produits déjà signalés, c'est-à-dire ceux des échantillons E1 à E8, E'1 à E'8 et E''1 à E''8, en supprimant les produits apparaissant plusieurs fois (on ne garde qu'un fromage de même code). Tableau 19, nous constatons que la variance et l'écart type d'erreur est plus faible pour la combinaison linéaire OWA-MAUT que pour ces deux modèles de notation seuls.

|  | OWA V=0,1 | OWA V=0,50 | OWA V=1 | MAUT V=0,1 | MAUT V=0,5 | MAUT V=1 | C.L. V=0,1 | C.L. V=0,50 | C.L. V=1 |
|---|---|---|---|---|---|---|---|---|---|
| Var. | 1,9909 | 0,7865 | 1,1851 | 0,3752 | 1,3749 | 0,8024 | 0,2189 | 0,7129 | 0,6482 |
| Ec. t | 1,4110 | 0,8868 | 1,0886 | 0,6125 | 1,1725 | 0,8957 | 0,4679 | 0,8443 | 0,8051 |

Tableau 19 : Variance d'écart avec les notes du décideur virtuel et écarts types

La variance est calculée comme la somme des carrés des écarts entre les notes du décideur virtuel et celles du modèle concerné, divisé par 28-1=27. Si nous comparons les résultats d'OWA seul et de MAUT seul, nous constatons que MAUT modélise mieux les notes du décideur que OWA, sauf pour une variance d'erreur de 0,5. Ce résultat curieux avait déjà été noté. Cependant, la combinaison linéaire OWA-MAUT est meilleure pour une variance de 0,1 que les deux modèles séparément.

## 8 Modélisation des notes par l'intégrale de CHOQUET

L'intégrale discrète de CHOQUET est souvent utilisée en tant que modèle de préférence (JACQUET-LAGRÈZE 1977) (DUBOIS et al. 1986) et aussi comme aide à la décision multicritère (KAYMAK et al. 1998) (WANG et al. 2003) (BEN-ARIEH 2005). De nombreuses études ont porté sur l'intégrale discrète de CHOQUET (MUROFUSHI et al. 1991) (MARICHAL 2002) avec des applications en classification (GRABISCH et al. 1994), en reconnaissance d'image (GRABISCH 1995), en reconnaissance de défauts (SCHMITT et al. 2008b), en reconnaissance de symboles graphiques (WENDLING et al. 2008a), en théorie du choix social, en description sémantique (CAMARGO et al. 2014), en fusion de données, en choix de tweets à proposer à un internaute (MOULAHI et al. 2015) et en webmarketing (TERRIEN 2018). Nous allons rappeler ce qu'est l'intégrale de CHOQUET discrète.

Soit m critères s'appliquant sur p produits. Chaque critère est normalisé entre 0 et 1 sous forme d'utilités $u_j(i)$ où $j \in \{1, \cdots m\}$ et $i \in \{1, \cdots p\}$. Pour un produit donné les valeurs des utilités sont classées en performances **croissantes** :

$$u_{\rho(0)}(i) = 0 \leq u_{\rho(1)}(i) \leq u_{\rho(2)}(i) \leq \cdots \leq u_{\rho(m)}(i) \leq u_{\rho(m+1)}(i) = 1 \qquad (13)$$

Où $j = \rho(k)$ est une fonction discrète qui donne l'indice j du critère dont l'utilité est de rang k dans le classement croissant des utilités du produit i. Donc $\rho(1)$ est le numéro du critère ayant la plus faible performance, et $\rho(m)$ est le numéro du critère de plus haute performance. Notons qu'il existe un lien entre la fonction $\rho(k)$ et $\sigma(j)$ présenté en section 3 pour OWA. La différence est que pour OWA les valeurs de critère étaient classées de façon **décroissante**, et donc que $\rho(k) = \sigma(m + 1 - k)$. De plus, les fonctions $\rho$ et $\sigma$ devraient être affectée de l'indice $i$ puisqu'elles changent d'un fromage à un autre, les classements en performance des critères changent d'un fromage à un autre. Nous ne le ferons pas pour ne pas alourdir la notation. Alors l'intégrale discrète de CHOQUET s'écrit :

$$C_\mu(i) = \sum_{k=1}^{m} \mu_{\rho(k)\cdots\rho(m)} \left( u_{\rho(k)}(i) - u_{\rho(k-1)}(i) \right) \qquad (14)$$

Les termes $\mu_{\rho(k)\cdots\rho(m)}$ sont les capacités de CHOQUET. Lorsqu'elles n'ont qu'un seul indice ($\mu_{\rho(m)}$) elles correspondent à un poids associé à un critère qui s'active si ce critère a la performance la plus élevée, lorsqu'elles ont plusieurs indices elles correspondent à des interactions entre critères. Dans ce dernier cas l'ordre d'écriture des indices n'a pas d'importance, on peut les réécrire dans un ordre croissant. Mais, pour l'ensemble de tous les critères (ou plutôt de toutes les performances) :

$$\mu_{\rho(1)\cdots\rho(m)} = 1 \qquad (15)$$

L'intégrale de CHOQUET peut aussi s'écrire :

$$C_\mu(i) = \sum_{k=1}^{m-1} u_{\rho(k)}(i) \left( \mu_{\rho(k)\cdots\rho(m)} - \mu_{\rho(k+1)\cdots\rho(m)} \right) + u_{\rho(m)} \mu_{\rho(m)} \qquad (16)$$

Où $u_{\rho(m)}\mu_{\rho(m)}$ est l'utilité de plus haute performance multipliée par son poids.

Les méthodes OWA et MAUT sont des cas particuliers de l'intégrale de CHOQUET discrète. En effet, nous avons pour OWA :

$$score_{OWA}(i) = \sum_{k=1}^{m} u_{\rho(k)}(i) w_k \qquad (17)$$

Ceci impose $\mu_{\rho(m)} = w_m$ et $\mu_{\rho(k)\cdots\rho(m)} = \mu_{\rho(k+1)\cdots\rho(m)} + w_k$, c'est-à-dire :

$$\mu_{\rho(k)\cdots\rho(m)} = \sum_{j=k}^{m} w_j \qquad (18)$$

| $\mu_1$ | 0,4418882 | $\mu_{13}$ | 0,7176442 | $\mu_{123}$ | 0,85139229 |
| --- | --- | --- | --- | --- | --- |
| $\mu_2$ | 0,4418882 | $\mu_{14}$ | 0,7176442 | $\mu_{124}$ | 0,85139229 |
| $\mu_3$ | 0,4418882 | $\mu_{23}$ | 0,7176442 | $\mu_{134}$ | 0,85139229 |
| $\mu_4$ | 0,4418882 | $\mu_{24}$ | 0,7176442 | $\mu_{234}$ | 0,85139229 |
| $\mu_{12}$ | 0,7176442 | $\mu_{34}$ | 0,7176442 | $\mu_{1234}$ | 1 |

Tableau 20 : Capacités de Choquet correspondant à OWA

Nous pouvons appliquer ceci au résultat du tableau 11, tout au moins à celui ayant la plus faible moyenne quadratique d'écart entre les poids du décideur virtuel et ceux identifiés dans le tableau 20. De même, MAUT s'écrit :

$$score_{MAUT}(i) = \sum_{k=1}^{m} u_{\rho(k)}(i) w_{\rho(k)} \qquad (19)$$

Ce qui impose $\mu_{\rho(m)} = w_{\rho(m)}$ et $\mu_{\rho(k)\cdots\rho(m)} = \sum_{j=k}^{m} w_{\rho(j)}$. Nous appliquons cela aux poids identifiés tableau 14 pour une variance d'erreur de notes égale à 1 (poids peu dispersés).

| $\mu_1$ | 0,24667935 | $\mu_{13}$ | 0,49256588 | $\mu_{123}$ | 0,73679183 |
| --- | --- | --- | --- | --- | --- |
| $\mu_2$ | 0,24422595 | $\mu_{14}$ | 0,50988751 | $\mu_{124}$ | 0,75411347 |
| $\mu_3$ | 0,24588653 | $\mu_{23}$ | 0,49011249 | $\mu_{134}$ | 0,75577405 |
| $\mu_4$ | 0,26320817 | $\mu_{24}$ | 0,50743412 | $\mu_{234}$ | 0,75332065 |
| $\mu_{12}$ | 0,4909053 | $\mu_{34}$ | 0,5090947 | $\mu_{1234}$ | 1 |

Tableau 21 : Capacités de Choquet correspondant à MAUT

En pratique, nous devons identifier les paramètres de l'intégrale de CHOQUET à partir des notes données par le décideur virtuel, nous devons donc poser :

$$\overline{note}_C(i) = aC_\mu(i) + b = b + \sum_{k=1}^{m} a\mu_{\rho(k)\cdots\rho(m)} \left( u_{\rho(k)}(i) - u_{\rho(k-1)}(i) \right) \qquad (20)$$

Les paramètres à identifier sont donc $b$ et tous les $a\mu_{\rho(k)\cdots\rho(m)}$ $\forall k \in \{1, \cdots m\}$. Compte tenu de l'équation (15) nous avons $a\mu_{\rho(1)\cdots\rho(m)} = a$. Les paramètres de l'intégrale de Choquet sont :

- Les 4 $\mu_{\rho(m)}$ correspondants aux 4 critères : $\mu_1, \mu_2, \mu_3$ et $\mu_4$,
- Les 6 interactions entre 2 critères : $\mu_{12}, \mu_{13}, \mu_{14}, \mu_{23}, \mu_{24}, \mu_{34}$,
- Les 4 interactions entre 3 critères : $\mu_{123}, \mu_{124}, \mu_{134}, \mu_{234}$,
- Et $\mu_{\rho(1)\cdots\rho(m)} = 1$ qui permet d'obtenir $a$.

Rappelons que pour les interactions l'ordre des indices de critères n'a pas d'importance. Nous avons donc 14+2=16 paramètres à identifier. Il faut donc sélectionner au moins 16 fromages parmi les 47, avec une difficulté : pour identifier $\mu_1, \mu_2, \mu_3$ et $\mu_4$, il faut que chaque critère soit au moins une fois la meilleure performance selon l'équation (16).

Il existe une approche de réseau itérative basée sur Choquet développée par (SCHMITT et al. 2008a) afin d'obtenir le modèle de décision le plus précis (SCHMITT et al. 2008b) (WENDLING et al. 2008b). Nous préfèrerions appliquer la méthode exposée section 3 par l'équation (6). Le problème est de construire la matrice $\bar{\bar{X}}$ qui, ici, est très creuse (au plus 5 éléments non nuls par ligne de 16 éléments). Il convient donc de choisir une matrice n'ayant aucune colonne nulle si elle est carrée, afin qu'elle puisse être inversible. Afin de voir la faisabilité de cette méthode, le tableau 22 présente le nombre de fois que l'on trouve des valeurs non nulles du facteur de chaque paramètre dans $\bar{\bar{X}}$ sur les 47 fromages.

Nous constatons que s'il est facile d'obtenir de bonnes estimations des paramètres a et b, il n'en est pas de même pour $\mu_{23}$, $\mu_{24}$ et même $\mu_{234}$. Pour obtenir au moins un résultat il faudrait faire l'identification paramétrique à partir des 47 fromages dont nous disposons, sans certitude concernant la qualité du résultat.

| $\mu_1$ | 22 | $\mu_{13}$ | 15 | $\mu_{123}$ | 14 |
|---|---|---|---|---|---|
| $\mu_2$ | 7 | $\mu_{14}$ | 12 | $\mu_{124}$ | 9 |
| $\mu_3$ | 8 | $\mu_{23}$ | 2 | $\mu_{134}$ | 20 |
| $\mu_4$ | 10 | $\mu_{24}$ | 2 | $\mu_{234}$ | 4 |
| $\mu_{12}$ | 10 | $\mu_{34}$ | 6 | A et b | 47 |

Tableau 22 : Nombre de fois que le facteur devant chaque capacité de Choquet est non nul

En fait, nous préférons simplifier le problème en remarquant que les équations (18) pour OWA et après l'équation (19) pour MAUT s'écrivent toutes les deux de la même façon :

$$\mu_{\rho(k)\cdots\rho(m)} = \sum_{j=k}^{m} \mu_{\rho(j)} \qquad (21)$$

Ainsi, cette équation (21) s'applique aussi à la combinaison linéaire de OWA et de MAUT. Nous l'étendrons à notre cas afin de réduire le nombre d'inconnues à identifier. En effet, avec l'équation (21) le score de l'intégrale de Choquet ainsi simplifiée est dans notre cas :

$$\left. \begin{array}{l} C_\mu(i) = \mu_{\rho(4)}\left(u_{\rho(4)}(i) - u_{\rho(3)}(i)\right) + \left(\mu_{\rho(4)} + \mu_{\rho(3)}\right)\left(u_{\rho(3)}(i) - u_{\rho(2)}(i)\right) + \\ \left(\mu_{\rho(4)} + \mu_{\rho(3)} + \mu_{\rho(2)}\right)\left(u_{\rho(2)}(i) - u_{\rho(1)}(i)\right) + u_{\rho(1)}(i) \end{array} \right\} \qquad (22)$$

Ce qui donne la note correspondante :

$$\left. \begin{array}{l} \overline{note}_C = b + a\mu_{\rho(4)}\left(u_{\rho(4)}(i) - u_{\rho(1)}(i)\right) + a\mu_{\rho(3)}\left(u_{\rho(3)}(i) - u_{\rho(1)}(i)\right) \\ + a\mu_{\rho(2)}\left(u_{\rho(2)}(i) - u_{\rho(1)}(i)\right) + au_{\rho(1)}(i) \end{array} \right\} \qquad (23)$$

Ainsi les paramètres sont au nombre de 4+2=6, $\vec{\theta}^T = (b \quad a \quad a\mu_1 \quad a\mu_2 \quad a\mu_3 \quad a\mu_4)$, mais admettant la contrainte due aux équations (15) et (21) :

$$\left. \begin{array}{l} \mu_{\rho(1)\cdots\rho(m)} = 1 = \sum_{j=1}^{m} \mu_{\rho(j)} = \sum_{j=1}^{m} \mu_j \Rightarrow \sum_{j=1}^{m} a\mu_j = a \\ (0 \quad -1 \quad 1 \quad 1 \quad 1 \quad 1)\vec{\theta} = 0 = \vec{z}^T\vec{\theta} \end{array} \right\} \qquad (24)$$

Nous devons minimiser $E$ de l'équation (5) sous la contrainte $\vec{z}^T\vec{\theta} = 0$, avec $\vec{z}^T = (0 \quad -1 \quad 1 \quad 1 \quad 1 \quad 1)$, c'est-à-dire minimiser :

$$\left.\begin{array}{c} E^+ = (\vec{\Sigma} - \vec{S})^T(\vec{\Sigma} - \vec{S}) + \lambda \vec{z}^T\vec{\theta} = (\bar{\bar{X}}\vec{\theta} - \vec{S})^T(\bar{\bar{X}}\vec{\theta} - \vec{S}) + \lambda \vec{z}^T\vec{\theta} \\ E = \vec{\theta}^T\bar{\bar{X}}^T\bar{\bar{X}}\vec{\theta} - 2\vec{S}^T\bar{\bar{X}}\vec{\theta} + \vec{S}^T\vec{S} + \lambda \vec{z}^T\vec{\theta} \\ \left(\frac{\partial E}{\partial \vec{\theta}}\right)^T = 2\bar{\bar{X}}^T\bar{\bar{X}}\vec{\theta} - 2\bar{\bar{X}}^T\vec{S} + \lambda \vec{z} = 0 \\ \frac{\partial E}{\partial \lambda} = \vec{z}^T\vec{\theta} = 0 \end{array}\right\} \quad (25)$$

Où $\lambda$ est un multiplicateur de LAGRANGE. La première des dérivées de (25) donne :

$$\hat{\vec{\theta}} = (\bar{\bar{X}}^T\bar{\bar{X}})^{-1}\left(\bar{\bar{X}}^T\vec{S} - \frac{\lambda}{2}\vec{z}\right) \quad (26)$$

La deuxième dérivée de (25) conduit à :

$$\left.\begin{array}{c} \vec{z}^T(\bar{\bar{X}}^T\bar{\bar{X}})^{-1}\left(\bar{\bar{X}}^T\vec{S} - \frac{\lambda}{2}\vec{z}\right) = 0 \\ \frac{\lambda}{2} = \frac{\vec{z}^T(\bar{\bar{X}}^T\bar{\bar{X}})^{-1}\bar{\bar{X}}^T\vec{S}}{\vec{z}^T(\bar{\bar{X}}^T\bar{\bar{X}})^{-1}\vec{z}} \end{array}\right\} \quad (27)$$

Ce qui donne finalement :

$$\left.\begin{array}{c} \hat{\vec{\theta}} = (\bar{\bar{X}}^T\bar{\bar{X}})^{-1}\left(\bar{\bar{X}}^T\vec{S} - \frac{\vec{z}^T(\bar{\bar{X}}^T\bar{\bar{X}})^{-1}\bar{\bar{X}}^T\vec{S}}{\vec{z}^T(\bar{\bar{X}}^T\bar{\bar{X}})^{-1}\vec{z}}\vec{z}\right) = (\bar{\bar{X}}^T\bar{\bar{X}})^{-1}\bar{\bar{X}}^T\vec{S} - \frac{\left((\bar{\bar{X}}^T\bar{\bar{X}})^{-1}\vec{z}\right)\vec{z}^T\left((\bar{\bar{X}}^T\bar{\bar{X}})^{-1}\bar{\bar{X}}^T\vec{S}\right)}{\vec{z}^T\left((\bar{\bar{X}}^T\bar{\bar{X}})^{-1}\vec{z}\right)} \\ \hat{\vec{\theta}} = (\bar{\bar{X}}^T\bar{\bar{X}})^{-1}\bar{\bar{X}}^T\vec{S} - \frac{\vec{z}^T\left((\bar{\bar{X}}^T\bar{\bar{X}})^{-1}\bar{\bar{X}}^T\vec{S}\right)}{\vec{z}^T\left((\bar{\bar{X}}^T\bar{\bar{X}})^{-1}\vec{z}\right)}(\bar{\bar{X}}^T\bar{\bar{X}})^{-1}\vec{z} = \hat{\vec{\theta}}^0 - \frac{\vec{z}^T\hat{\vec{\theta}}^0}{\vec{z}^T\vec{z}^0}\vec{z}^0 \end{array}\right\} \quad (28)$$

Avec $\hat{\vec{\theta}}^0 = (\bar{\bar{X}}^T\bar{\bar{X}})^{-1}\bar{\bar{X}}^T\vec{S}$ et $\vec{z}^0 = (\bar{\bar{X}}^T\bar{\bar{X}})^{-1}\vec{z}$.

L'identification paramétrique est réalisée à l'aide de l'échantillon de 9 fromages formé de E' et de E''4, déjà utilisé pour la combinaison linéaire d'OWA et de MAUT. Les paramètres a et b sont :

| CHOQUET | Variance 0.1 | Variance 0.5 | Variance 1 |
|---|---|---|---|
| $b$ | -10,2856958 | -9,15774527 | -9,67486721 |
| $a$ | 25,7472056 | 24,0233479 | 24,8819919 |

Tableau 23 : Paramètres a et b de l'intégrale de Choquet

Les valeurs de a et b présentées tableau 23 sont plus faibles de celles du décideur virtuel mais restent proches de celles obtenues avec MAUT. Les valeurs de capacité de CHOQUET sont dans le tableau 24 :

| $\mu_1$ | 0,30762376 | $\mu_{13}$ | 0,52424619 | $\mu_{123}$ | 0,77723344 |
|---|---|---|---|---|---|
| $\mu_2$ | 0,25298725 | $\mu_{14}$ | 0,53039032 | $\mu_{124}$ | 0,78337758 |
| $\mu_3$ | 0,21662242 | $\mu_{23}$ | 0,46960968 | $\mu_{134}$ | 0,74701275 |
| $\mu_4$ | 0,22276656 | $\mu_{24}$ | 0,47575381 | $\mu_{234}$ | 0,69237624 |
| $\mu_{12}$ | 0,56061102 | $\mu_{34}$ | 0,43938898 | $\mu_{1234}$ | 1 |

Tableau 24 : Capacités de Choquet identifiées à partir des notes à variance d'erreur égale à 0,1

| $\mu_1$ | 0,18170731 | $\mu_{13}$ | 0,45989295 | $\mu_{123}$ | 0,76838102 |
|---|---|---|---|---|---|
| $\mu_2$ | 0,30848806 | $\mu_{14}$ | 0,41332629 | $\mu_{124}$ | 0,72181436 |
| $\mu_3$ | 0,27818564 | $\mu_{23}$ | 0,58667371 | $\mu_{134}$ | 0,69151194 |
| $\mu_4$ | 0,23161898 | $\mu_{24}$ | 0,54010705 | $\mu_{234}$ | 0,81829269 |
| $\mu_{12}$ | 0,49019537 | $\mu_{34}$ | 0,50980463 | $\mu_{1234}$ | 1 |

Tableau 25 : Capacités de Choquet identifiées à partir des notes à variance d'erreur égale à 0,5

Dans le cas d'une variance d'erreur de notes égale à 0,5 les valeurs de capacité sont données tableau 25. De même, pour une variance d'erreur des notes égale à 1, les valeurs de capacité de Choquet sont :

| $\mu_1$ | 0,25405211 | $\mu_{13}$ | 0,47305334 | $\mu_{123}$ | 0,75232717 |
|---|---|---|---|---|---|
| $\mu_2$ | 0,27927383 | $\mu_{14}$ | 0,50172494 | $\mu_{124}$ | 0,78099877 |
| $\mu_3$ | 0,21900123 | $\mu_{23}$ | 0,49827506 | $\mu_{134}$ | 0,72072617 |
| $\mu_4$ | 0,24767283 | $\mu_{24}$ | 0,52694666 | $\mu_{234}$ | 0,74594789 |
| $\mu_{12}$ | 0,53332594 | $\mu_{34}$ | 0,46667406 | $\mu_{1234}$ | 1 |

Tableau 26 : Capacités de Choquet identifiées à partir des notes à variance d'erreur égale à 1

Ces résultats restent relativement proches de ceux présentés tableau 21 pour MAUT. Il est donc important de valider l'intégrale de Choquet obtenue à l'aide de l'ensemble de 28 produits utilisés pour la combinaison linéaire OWA-MAUT. Il s'agit de calculer la variance d'écart entre les notes données par le décideur et celle obtenue avec l'intégrale de Choquet, et ce pour les différentes variances d'erreur de notation du décideur virtuel. Ces variances et l'écart type correspondant sont données tableau 27. Ces résultats montrent des valeurs de variance et d'écart type supérieures à celles de la combinaison linéaire OWA-MAUT. Par contre, elles sont proches des variances d'écart des notes de MAUT, et meilleures que deux sur trois valeurs correspondantes de OWA (les cas de variance d'erreur de notation 0,1 et 1).

| CHOQUET | V=0,1 | V=0.5 | V=1 |
|---|---|---|---|
| Variance | 0,43767599 | 1,33787208 | 0,69017648 |
| Ecart type | 0,66157085 | 1,1566642 | 0,830768608 |

Tableau 27 : Variance et écart type de l'écart des notes du décideur et de l'intégrale de Choquet

## 9 Conclusions et perspectives

L'objectif du présent papier est de prendre la suite de (RENAUD et al. 2008), plus particulièrement en montrant que l'utilisation d'une technique de construction de plans d'expériences comme la D-optimalité est capable d'améliorer la précision de l'identification paramétrique appliquée à la modélisation décisionnelle. Un autre objectif est de comparer les modèles OWA (Ordered Weighted Average) et MAUT (Multi Attribute Utility Theory) avec une combinaison linéaire OWA-MAUT et une intégrale discrète de CHOQUET.

Afin de réaliser l'identification paramétrique à partir de notes, un décideur virtuel a été défini, le décideur réel n'étant pas disponible sur un temps suffisamment long. Les notes ont été attribuées à 47 fromages de munster produits en trois mois et goûtés par des spécialistes, les utilités de quatre critères ayant été calculées à partir de leurs appréciations. Chaque note a été bruitée artificiellement avec trois variances d'erreur de notation afin de rendre compte des erreurs humaines potentielles, et d'étudier l'influence de l'amplitude de cette erreur. L'identification paramétrique effectuée à partir de classement a été possible en déduisant les classements indispensables des notes.

La section 4 est consacrée à l'identification paramétrique des poids de OWA à partir des notes. Le calcul est fait pour 8 échantillons de 5 fromages. Sur ces huit échantillons, seuls 1 à 3 échantillons sont pertinents au regard des variances d'erreur de notation de 0,1 à 1, soit entre un huitième et un tiers des échantillons. Ceci montre la difficulté de trouver un échantillon pertinent, permettant une identification précise. La section 5 présente l'identification paramétrique des poids de OWA à partir des classements. Les résultats sont presque identiques à l'identification à partir des notes. Là encore les échantillons utilisés semblent manquer de pertinence. L'échantillon E1 semble être le plus robuste, il propose pour chaque profil de décideur, un ensemble de poids positifs. L'ensemble des poids est très différent entre eux. Ils passent de 0,34 à 0,6 pour le premier poids ; de 0,02 à 0,72 pour le deuxième poids ; de 0,09 à 0,26 pour le troisième poids, et de 0,03 à 0,27 pour le quatrième poids.

Ainsi, à partir de ces 8 échantillons nous appliquons une méthode dite D-optimale pour obtenir 8 nouveaux échantillons de 5 fromages, dont la matrice d'information de FISCHER ait le déterminant le plus élevé possible. Quel que soit la variance d'erreur de notation, les résultats de l'identification paramétrique à partir des notes sont bien meilleurs que ceux à partir des classements. Point important, les résultats de l'identification paramétrique à partir des notes pour le meilleur échantillon obtenu par D-optimalité est considérablement meilleur que pour le meilleur des 8 échantillons initiaux. Concernant le classement des échantillons initiaux pour OWA, l'ordre proposé par les notes était le même que la proposition du décideur réel (RENAUD et al. 2008). Pour le meilleur échantillon obtenu par D-optimalité, l'identification paramétrique à partir du classement n'améliore pas la précision des poids (estimée par la moyenne quadratique d'écart entre les poids du décideur virtuel et ceux identifiés).

Comme le choix d'un modèle de préférence devant représenter un décideur réel n'est pas une tache facile, une erreur est possible. Pour rendre compte d'une telle erreur, nous avons testé le modèle MAUT. Le résultat de l'identification paramétrique à partir de notes et à partir de classement est moins bon pour MAUT que pour OWA. En effet, sur les 8 échantillons deux ne donne aucun jeu de poids valides par OWA et le double par MAUT. Globalement OWA fournit 20 jeux de poids valides sur 6x8=48, alors que MAUT n'en donne que 13. Ceci semble normal puisque le décideur virtuel est basé sur OWA. L'amélioration des échantillons par D-optimalité donne le meilleur échantillon pour MAUT, E''4, qui donne des jeux de poids cohérents par identification à partir des notes quel que soit la variance d'erreur de notation. Cependant, bien que les notes donnent le même classement quel que soit la variance d'erreur de notation, le jeu de poids obtenu à partir du classement est invalide. D'un point de vue pratique, nous pensons que l'identification paramétrique par les notes doit être privilégiée par rapport à celle par classement, même si cela donne plus de travail au décideur réel.

Ce résultat amélioré peut l'être encore plus si nous réalisons une combinaison linéaire OWA-MAUT. C'est ce qui nous a donné l'idée d'essayer le modèle basé sur l'intégrale discrète de CHOQUET. En effet, OWA et MAUT sont des cas particuliers d'intégrale discrète de CHOQUET. L'identification paramétrique semble indiquée pour estimer les 16 paramètres du modèle, cependant le tableau 22 montre la difficulté de calculer certains paramètres d'interaction. C'est pourquoi nous avons choisi de réduire le nombre de paramètres en calculant ceux d'interaction à partir des quatre premières capacités, celles ayant un comportement de poids. Les résultats de simulation des notes est bon, car meilleurs que ceux de OWA et MAUT seuls, cependant ces résultats sont un peu moins bons que la combinaison linéaire OWA-MAUT.

En perspective, Il devrait être possible d'améliorer un peu les résultats de l'intégrale de CHOQUET en prévision de notes grâce à la D-optimalité. Cependant, il faudrait aussi libérer les paramètres d'interaction des contraintes que nous avons imposées. Il faudrait donc essayer la technique FRIFS (SCHMITT et al. 2008a) (SCHMITT et al. 2008b) pour améliorer les capacités de CHOQUET obtenues tableaux 24, 25 et 26.

**Références**